\documentclass[prb,twocolumn,showpacs]{revtex4}
\usepackage{graphicx,epsf}
\usepackage{bm, bbm,mathbbol}      

\newcommand{\bq}{{\bf q}}
\newcommand{\bk}{{\bf k}}

\newcommand{\bK}{{\bf K}}
\newcommand{\bQ}{{\bf Q}}

\newcommand{\beq}{\begin{equation}}
\newcommand{\beqn}{\begin{eqnarray}}
\newcommand{\eeq}{\end{equation}}
\newcommand{\eeqn}{\end{eqnarray}}

\usepackage{color}

\begin{document}

\def\tende#1{\,\vtop{\ialign{##\crcr\rightarrowfill\crcr
\noalign{\kern-1pt\nointerlineskip}
\hskip3.pt${\scriptstyle #1}$\hskip3.pt\crcr}}\,}

\title{Topologically Protected Zero Modes in Twisted Bilayer Graphene}
\author{R. de Gail$^{1}$, M. O. Goerbig$^{1}$, F. Guinea$^{2}$, G. Montambaux$^{1}$, and A. H. Castro Neto$^{3,4}$}

\affiliation{
$^1$Laboratoire de Physique des Solides, CNRS UMR 8502, Univ. Paris-Sud, F-91405 Orsay cedex, France\\
$^2$Instituto de Ciencia de Materiales de Madrid (CSIC), Sor Juana In\'es de la Cruz 3, E-28049 Madrid, Spain.\\
$^3$Graphene Research Centre and Physics Department, National University of Singapore, 2 Science Drive 3, Singapore 117541\\
$^4$Department of Physics, Boston University, 590 Commonwealth Avenue, Boston, MA 02215, USA}

\begin{abstract}
We show that the twisted graphene bilayer can reveal unusual topological properties at low energies, as a consequence of a Dirac-point splitting. 
These features rely on a symmetry analysis of the electron hopping between the two layers of graphene and we derive a simplified effective low-energy Hamiltonian 
which captures the essential topological properties of twisted bilayer graphene. The corresponding Landau levels peculiarly reveal a degenerate 
zero-energy mode which cannot be lifted by strong magnetic fields.
\end{abstract}
\maketitle

\section{Introduction}

One of the most fascinating aspects of graphene is its band structure which can be fundamentally changed in several
different ways by modifying its lattice structure. This happens because the honeycomb lattice of monolayer graphene
has two independent sublattices and the electron, as it moves through the lattice, has to change its sublattice and
hence the character of its wavefunction.\cite{Revs} Thus, even small local changes in the lattice structure lead to the appearance
of gauge potentials which are associated with the phase of the electronic wavefunction in each sublattice. As a consequence, 
there is a one-to-one correspondence between graphene's structure with the topological features of the electronic states. 

Another amazing property of graphene is its honeycomb structure which yields an electronic low-energy effective theory
which is \textit{Lorentz-invariant} in two dimensions (2D), and thus corresponds to 2D Dirac fermions. This Lorentz
invariance is robust because the energy associated with sublattice coupling, that is, the inter-sublattice hopping energy
$t$ ($ \approx 3$ eV) is the dominant energy scale in the system. Lorentz invariance persists even when the lattice is
modified either by external forces (strain, shear, etc.),\cite{strain} by external fields, or by the addition of more 
layers.\cite{Lorentz} For instance,
AB-stacked bilayer graphene is described, at low energies by two sets of massive Lorentz invariant Dirac particles per valley and spin. 
In the simplest models where only nearest neighbor hoppings are taking into account, there is still 
an accidental degeneracy that makes the particle-like band of one flavor to be degenerate with the anti-particle-like (hole-like) 
band of the other flavor at the K (K') point in each valley. This degeneracy can be easily lifted by the application of a
perpendicular electric field that breaks the inversion symmetry in the system.\cite{Revs} Although the Lorentz invariance
is preserved, the wavefunction of the electrons at low energies is modified$-$whereas in monolayer graphene the Dirac fermions
carry a Berry phase $\pm \pi$, the Berry phase is $\pm 2\pi$ in AB-stacked bilayer graphene.\cite{MF06} At low energies, trigonal 
warping splits this ``double'' Dirac point into three with a Berry phase $\pi$ and an additional one with $-\pi$,\cite{MF06} 
a situation that persists for a translational mismatch between the layers.\cite{Son}

Twisted bilayer graphene, in which the two layers have a rotational mismatch described by an
angle $\theta$ as compared to the perfect AB stacking, is another example where lattice structure and wavefunction topology
are directly interconnected. In fact, from the experimental point of view, twisted graphene is more the rule than the exception. 
It naturally occurs at the surface of graphite,\cite{Rong,Pong} in graphene grown in the surface of SiC,\cite{Hass} or
graphene grown by chemical vapor deposition on metal substrates.\cite{Eva} As compared to monolayer graphene, each Dirac flavor 
is then split into two copies that are separated in reciprocal space by a wave vector $\Delta\bK$, as a function of $\theta$.\cite{Lopes}
Inter-layer hopping results in a renormalization of the Fermi velocity\cite{Heer,Schmidt,Lopes,Shallcross} as well as 
in a van Hove singularity at relatively low energy as compared to that in monolayer graphene.\cite{Lopes,Eva} Once again,
Lorentz invariance is preserved at low energies but the nature of the electronic wavefunctions is modified in a profound way. 

In this paper, we investigate the topological aspects of the band structure of twisted graphene bilayers at low energies in the continuum
approximation, for small angle mismatches as compared to perfect AB stacking. We identify possible topological classes that describe band 
inversion symmetry. These topological classes determine the relative Berry phase between the two copies of Dirac particles, which 
have either the same or opposite Berry phase. If the two Dirac cones are related by time-reversal symmetry, such as the 
$K$ and $K'$ points in monolayer graphene, the Berry phases are naturally opposite. In twisted graphene bilayers, however, the two Dirac points emanating from different layers are not time-related, and the symmetry of the inter-layer hopping term enforces 
the Berry phases to be identical. We show that this feature yields a topologically
protected zero-energy Landau level, in contrast to the former case. The scenario may be tested in 
quantum Hall measurements. 

The paper is organized as follows. In Sec. \ref{sec:Mod4}, we discuss the model of twisted bilayer graphene and the symmetry 
properties of their bands (Sec. \ref{sec:sym}) that are fixed by the form of the inter-layer hopping. Furthermore, we present an
effective two-band model (Sec. \ref{sec:Mod2}) that displays the same topological low-energy properties as the original four-band model.
Section \ref{sec:LL} is devoted to the discussion of the Landau-level spectrum in twisted bilayer graphene, in the perspective of
possible quantum-Hall measurements.


\begin{figure}[t]
{\label{bla}\includegraphics[width=0.18\textwidth]{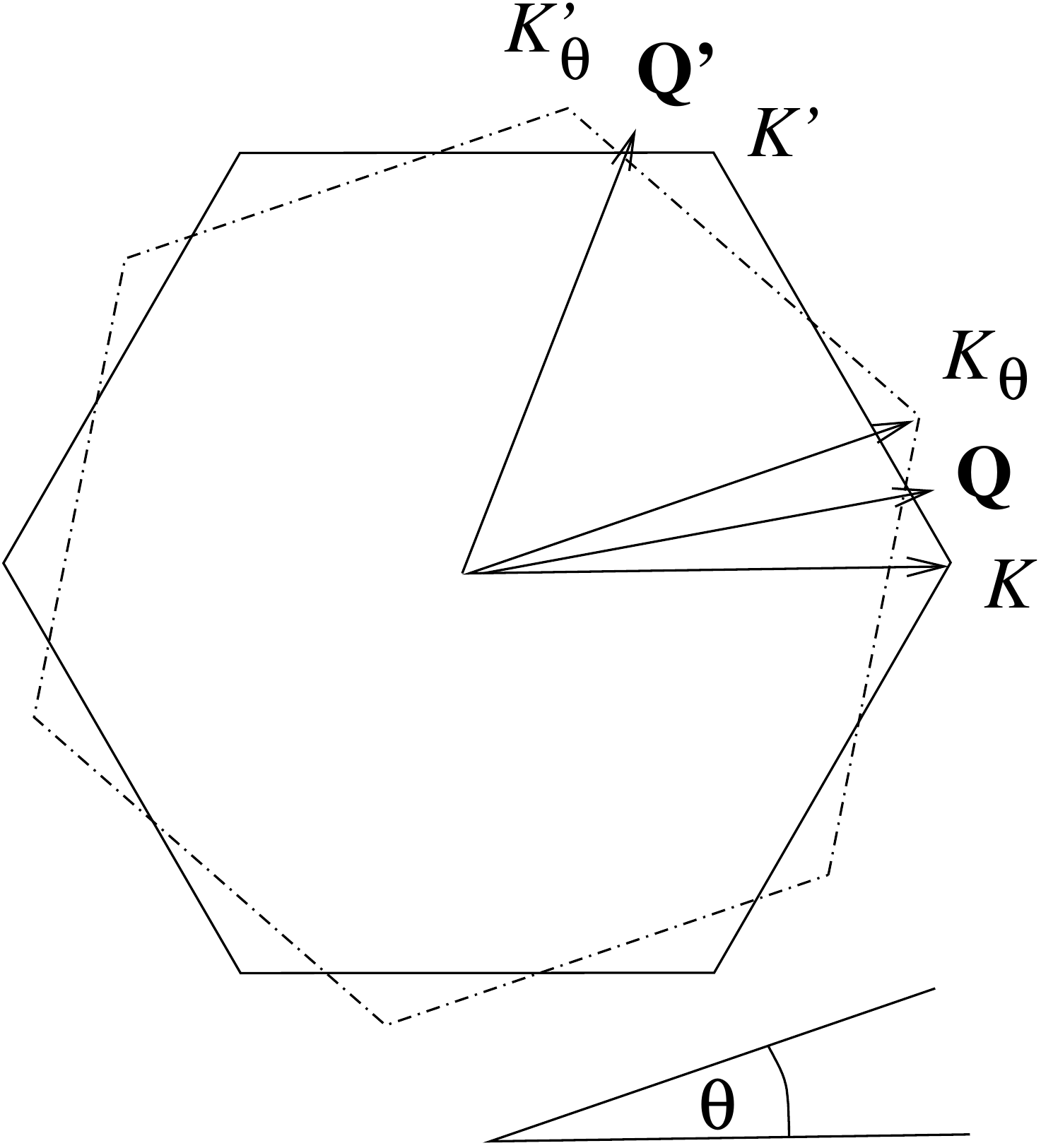}} 
\caption{First Brillouin zone for twisted bilayer graphene. The 1BZ of the upper layer (dashed hexagon) is rotated by an 
angle $\theta$ with respect to that of the lower layer (full hexagon). The corners, where Dirac points occur, are labeled by
$K_{\theta}^{(\prime)}$ and $K^{(\prime)}$, respectively. 
}
\label{fig01}
\end{figure}

\section{Model of twisted bilayer graphene}
\label{sec:Mod4}

If we neglect, for the moment, hopping
between atoms in different layers, the electronic properties of twisted bilayer graphene are 
described by two copies of the Hamiltonian for monolayer graphene (we use units with $\hbar=1$):
\beq\label{eq:ham0}
H_0(\bk)\equiv v_F \left(\begin{array}{cc} 0 & k^*\\ k & 0
                      \end{array}\right),
\eeq
where $v_F$ is the Fermi velocity and $k=k_x+i k_y$ is the 2D wave-vector relative to the K (K') points of the rotated
layers (see Fig.\ref{fig01}). For a twist (rotation) angle $\theta\neq 0$, each of the two inequivalent Dirac points, which reside at the corners of the first BZ $K$ and $K'$, 
are split into two, separated by a wave vector $\Delta \bK=\bK -\bK_{\theta}$, where $\bK_{(\theta)}$ 
is the position of the $K$ point in the lower (upper) layer and $-\bK_{(\theta)}$ the position of the 
points $K'$ and $K_{\theta}'$, respectively. 
Throughout this paper, we will work in the continuum limit around a single pair ($K$,$K_{\theta}$) and, therefore, neglect commensuration effects between the two layers which could 
enlarge the unit cell in position space and thus fold it back in reciprocal space. This procedure is mostly justified because the coupling between the pairs of Dirac cones\cite{Mele,MacDonald} is negligible.\cite{Shallcross}

The Hamiltonian describing the electronic properties of twisted bilayer graphene reads
\beq\label{eq:ham1}
H(\bk)=\left(\begin{array}{cc} H_0(\bk+\Delta \bK/2) & H_{\perp}\\ H_{\perp}^{\dagger} & H_0(\bk-\Delta\bK/2)
     \end{array}\right),
\eeq
where $H_{\perp}$ is the hopping matrix between the two layers.
Equation (\ref{eq:ham1}) refers to an expansion around the $\bQ$ point of Fig. \ref{fig01}. 
The analysis of the Moir\'e pattern formed by the twisted bilayer shows that, for small angle $\theta$, the hopping matrix
$H_{\perp}$ may have three different forms corresponding to the three main Fourier components.\cite{Lopes,MacDonald}
This leads to three different types of inter-layer hopping terms,
\beq
\label{eq:hopping}
H_{\perp}^0 \equiv  \tilde{t}_{\perp}
\left( 
\begin{array}{cc}
1 & 1 \\
1 & 1 \\
\end{array}
\right),
\quad
H_{\perp}^{\pm}  \equiv  \tilde{t}_{\perp}
\left( 
\begin{array}{cc}
e^{\mp i\phi} & 1 \\
e^{\pm i\phi} & e^{\mp i\phi} \\
\end{array}
\right),
\eeq
where $\phi=2\pi/3$ and $\tilde{t}_{\perp}$ is a hopping parameter which generally depends on $\theta$.\cite{Lopes,MacDonald,note}

\subsection{Symmetry of the bands and Berry phases}
\label{sec:sym}

In contrast to a lattice Hamiltonian that may be analyzed with the help of global symmetries, such as time reversal or lattice inversion, 
$H(\bk)$ in Eq. (\ref{eq:ham1}) is a continuum model in which the two Dirac points are no longer related by these symmetries. 
However, $H(\bk)$ and the inter-layer hopping term $H_{\perp}$ may be investigated via symmetries that involve directly 
the energy bands, such as rotation, mirror, and inversion symmetry. Whereas for $H_{\perp}=0$ the rotation and inversion symmetries
are respected, the latter are broken for non-zero inter-layer hopping, and we therefore restrict the discussion to the inversion
symmetry $\mathcal{I}$ of the bands. We emphasize that this inversion symmetry is defined with respect to the bands in reciprocal space, in contrast to a previous analysis,\cite{Paco} in which the more common definition of inversion symmetry with respect
to the lattice was used.

In the absence of any inter-layer hopping, the Hamiltonian (\ref{eq:ham1}) would be 
ambiguous since one could work equally well with $H_0^*$
in the second layer. As a consequence, there remain two possible representations of inversion (with different spinorial expressions) $\mathcal{I}_1$ and $\mathcal{I}_2$:
\beq
\label{eq:inv2}
\mathcal{I}_1:H(-\bk)=-H(\bk) \qquad {\rm or} \qquad \mathcal{I}_2: H(-\bk)=-H(\bk)^*.
\eeq
The minus signs in both transformations maps positive to negative energy states 
at opposite wave vectors and {\sl vice-versa}: $E(\bk)\rightarrow -E(-\bk)$, 
whereas the complex conjugation in $\mathcal{I}_2$ changes the relative phase between the spinor components and, hence, 
the Berry phase of a cone.

The phases of the two Dirac points are thus different for the two representations.
For the $\mathcal{I}_1$ case, the Berry phases at fixed energy
are opposite such that a merging transition of the two Dirac points (at $\theta=0$) corresponds to a vanishing total Berry phase and consequently to the possible opening 
of a band gap. This situation arises e.g. in the framework of the model discussed in Refs. \onlinecite{listDPmerg}, which describes
monolayer graphene under strong strain.\cite{strain}
In contrast to this rather well-known topological universality class, the transformation
$\mathcal{I}_2$ yields Berry phases that are the same at fixed energy $E$ and opposite to those at $-E$ and thus represents
a second universality class for Hamiltonians describing pairs of Dirac points. 
The transformation $\mathcal{I}_2$ may be represented in terms of a tensor product of Pauli matrices, 
$\mathcal{I}_2=\sigma_y^E\otimes \sigma_x^A$, where $\sigma_{x/y}^{A}$ and $\sigma_{x/y}^{E}$ describe the intra-layer
and the inter-layer spinorial spaces, respectively. 
The phases are eventually fixed by the symmetry of the interlayer hopping $H_{\perp}$, and the particular forms (\ref{eq:hopping})
happen to be invariant under $\mathcal{I}_2$.

Notice, however, that as a consequence of the 
inter-layer hopping terms (\ref{eq:hopping}), the two Dirac points are no longer exactly at the same energy, but we
neglect this energy shift here because it is associated with a very small energy scale $\sim 1$ meV. As a second-order perturbation, this indeed scales like $\tilde{t}_{\perp}^2/(v_F \Delta \bK)$ with $\tilde{t}_{\perp}$ of the order of $100$ meV and $v_F \Delta \bK \sim 1$ eV.\cite{Lopes}

\subsection{Effective two-band model}
\label{sec:Mod2}

In the opposite limit, $\tilde{t}_{\perp}\gg v_F\Delta K$, the model (\ref{eq:ham1}) may be reduced to an
effective two-band model, similarly to the case of a perfectly AB-stacked ($\theta=0$) graphene bilayer.\cite{MF06}
This is done in two steps. First, one replaces $H_{\perp}$ in Eq. (\ref{eq:ham1})
by a simplified inter-layer hopping term, 
\beq\label{eq2}
H_{\perp}^{\rm eff}=\tilde{t}_{\perp}\left(\begin{array}{cc} 
 0 & 0 \\
 1 & 0 
 \end{array}\right),
\eeq
in the limit where
$\tilde{t}_{\perp}\gg v_F\Delta K$, i.e. for small tilt angles. The inter-layer hopping term (\ref{eq2})
is reminiscent of the Bernal bilayer case. 
In spite of this simplification and the difference in the energy scales,
the resulting Hamiltonian may be viewed as a representative of the 
topological universality class that also includes the original four-band model.
We consider the eigenvectors of Eq. (\ref{eq2}) in terms of the 4-spinor basis $\left\{ \psi_A, \psi_B, \psi_{A'}, \psi_{B'} \right\}$ where $(A,B)$ belongs to the two sublattices of the first layer and $(A',B')$ to that of the second one.
With the particular form of Eq. (\ref{eq2}), the zero-energy sector is spanned
$\psi_A$ and $\psi_{B'}$, whereas $\psi_{B}$ and $\psi_{A'}$ are strongly hybridized by the inter-layer hopping term. 
Their symmetric and anti-symmetric combinations are the eigenstates at $-\tilde{t}_{\perp}$ and $\tilde{t}_{\perp}$, respectively.
In order to describe the electronic properties in the vicinity of $E=0$, one may therefore project the Hamiltonian onto the reduced $\left\{ \psi_A,\psi_{B'}\right\}$ basis and neglecting terms of the form $E \psi_{B/A'}$, which are a product of the energy
$E\sim 0$ and the small components $\psi_{B/A'}$. The eigenvalue problem reads
\beqn
\label{eq3a}
v_F (k+\Delta K/2)^*\psi_B  & = & E\psi_A\\
\label{eq3b}
v_F (k+\Delta K/2)\psi_A + \tilde{t}_{\perp}\psi_{A'} & = & E \psi_B \simeq 0 \\
\label{eq3c}
\tilde{t}_{\perp}\psi_{B} + v_F (k-\Delta K/2)^*\psi_{B'} & = & E\psi_{A'} \simeq 0 \\
\label{eq3d}
v_F (k-\Delta K/2)\psi_{A'} & = & E\psi_{B'}.
\eeqn
By rewriting Eqs. (\ref{eq3c}) and (\ref{eq3d}),
\beqn
-v_F/\tilde{t}_{\perp} (k+\Delta K/2)\psi_A  & =&   \psi_{A'} \\
-v_F/\tilde{t}_{\perp} (k-\Delta K/2)^*\psi_{B'}  & =&   \psi_{B},
\eeqn
and substituting them into Eqs. (\ref{eq3a}) and (\ref{eq3d}), one obtains the Schr\"odinger equation 
\beq
H^{\text{eff}}(\bk)\left(
\begin{array}{c}
\psi_A\\ \psi_{B'}
\end{array}
\right)
= E
\left(
\begin{array}{c}
\psi_A\\ \psi_{B'}
\end{array}
\right),
\eeq
in terms of the effective two-band Hamiltonian
\beq
\label{eq:hamKK}
H^{\text{eff}}(\bk)=-\frac{v_F^2}{\tilde{t}_{\perp}}
\left(
\begin{array}{cc}
0 & \left(k^*\right)^2 -\left(\Delta K^*\right/2)^2 \\
k^2-\left(\Delta K/2\right)^2 & 0 \\
\end{array}
\right),
\eeq
which is similar to that for the nematic transition of the interacting Bernal graphene bilayer.\cite{Vafek}

\begin{figure}[t]
\centering
{\label{bla1}\includegraphics[width=0.4\textwidth]{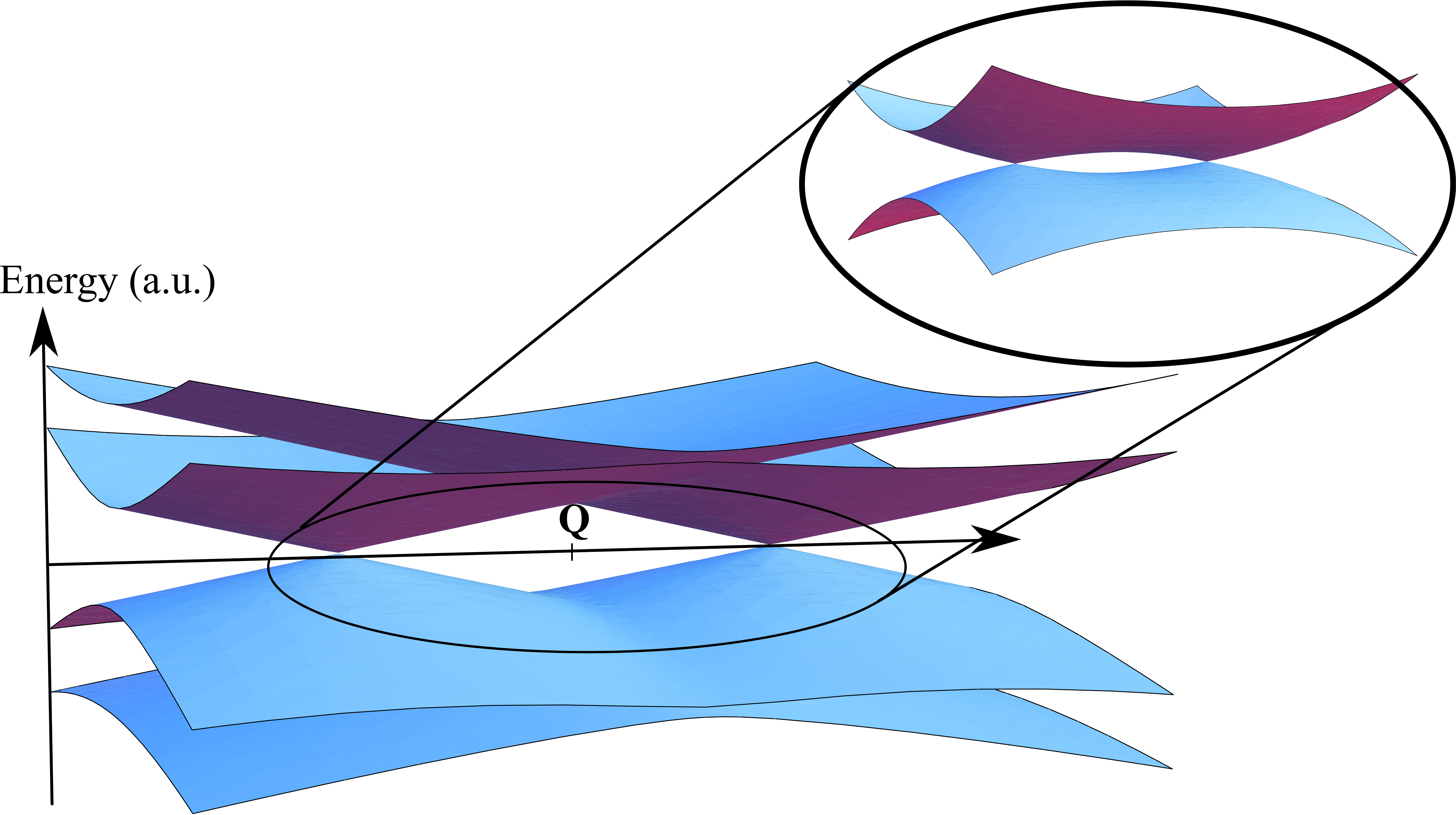}} 
\caption{(Color online) Generic band structure for the lowest four bands of the twisted bilayer graphene, around the $\bQ$ point, with only $H_{\perp}^0$ considered. The inset of the figure pictures the effective bands of the model Hamiltonian (\ref{eq:hamKK}). Both band structures are inversion-invariant with respect to the $\bQ$ point.}
\label{fig03}
\end{figure}

The generic form of the band structure obtained from the effective two-band model (\ref{eq:hamKK}) is depicted in the inset of Fig. \ref{fig03}.
One notices the two Dirac points originally situated at $\bK$ and $\bK_{\theta}$, separated by a wave vector $|\Delta \bK|\sim \theta/a$
in the first BZ, in terms of the intra-layer distance $a=0.142$ nm between neighboring carbon atoms. 
In order to see the linearity of the dispersion relation in the vicinity of these contact points, one can further expand the Hamiltonian 
(5) around $\pm \Delta \bK/2$, by defining $\bk = \bq \pm \Delta \bK/2$, with $|\bq| \ll |\Delta \bK|/2$. This expansion yields two
Dirac Hamiltonians 
\beq\label{eq:twoD}
\pm \frac{v_F^2 \Delta K}{\tilde{t}_{\perp}}
\left(
\begin{array}{cc}
0 & q^* \\
q & 0 
\end{array}
\right),
\eeq
with identical chirality for the two contact points. 

Furthermore, 
the bands have saddle points at $\bk=0$ between the two Dirac points. The effective
two-band Hamiltonian therefore captures the logarithmic van-Hove singularity 
in agreement with previous theoretical\cite{Lopes} and experimental studies.\cite{Li10}

The most salient features of the Hamiltonian (\ref{eq:hamKK}) are its chiral properties. 
In agreement with the above-mentioned general symmetry considerations, electrons at a fixed energy
at the $\bK$-point have the same chirality as those at the $\bK_{\theta}$ point, such that in both cases the electron
experiences the same Berry phase $\gamma=\pi$ (and $\gamma=-\pi$ at the points $\bK'$ and $\bK_{\theta}^{\prime}$)
on a closed orbit around one of the Dirac points. This is obvious in the low-energy expansion (\ref{eq:twoD}) around the
two Dirac points. As for AB-stacked bilayer graphene,
the Berry phase acquired on an orbit enclosing both Dirac points is then $2\gamma$, as one may also see from the limit
of vanishing twist angle ($\theta=0$) which reproduces the perfectly AB-stacked bilayer.\cite{MF06} Furthermore, it becomes apparent
from the form of the Hamiltonian (\ref{eq:hamKK}) that the merging of the Dirac points is not accompanied with a gap opening,
in contrast to the Hamiltonian discussed in Ref. \onlinecite{listDPmerg}, which desribes the same band structure in the semi-metallic 
phase but which belongs to another topological class, described by the symmetry $\mathcal{I}_1$.

\section{Landau levels of twisted bilayer graphene}
\label{sec:LL}

One of the most prominent consequences of these topological properties is the presence of a (doubly-degenerate) zero mode
that emerges in the presence of a quantizing magnetic field, in which case the Hamiltonian (\ref{eq:hamKK}) may be
written in terms of the usual ladder operators $a$ and $a^{\dagger}$, with $[a,a^{\dagger}]=1$:
\beq\label{eq:hamKK_B}
H_B=\omega_C \left(
\begin{array}{cc} 0 & a^{2} - \alpha^{*2} \\ a^{\dagger 2} - \alpha^{2} & 0 \end{array}
\right),
\eeq
where $\omega_C= 2v_F^2eB/\tilde{t}_{\perp}$ is the cyclotron frequency and  $\alpha\equiv \Delta K l_B/2\sqrt{2}$. 
As compared to the perfectly AB-stacked bilayer ($\alpha=0$), where one readily obtains the Landau level (LL) spectrum,\cite{MF06}
one notices that the additional
terms couple states only of the same parity. One obtains thus two classes of eigenstates $\psi_{2n}$ and $\psi_{2n+1}$
that may be written in the usual
harmonic-oscillator basis $|m\rangle$, with $a^{\dagger}a|m\rangle = m|m\rangle$:
\beq\label{eq:eigen}
\Psi_{2n(+1)} = \sum_{m=0}^{\infty}\left(\begin{array}{c} \phi_{2m(+1)}^1 \\ \phi_{2m(+1)}^2
\end{array}\right) |2m(+1)\rangle,
\eeq
where the components $\phi_{m}^1$ and $\phi_{m}^2$ are to be determined recursively. 

\subsection{Zero-energy levels}
\label{sec:0LL}

Before discussing the LL spectrum of Hamiltonian (\ref{eq:hamKK_B}), we investigate the zero-energy states, which may be obtained
analytically from the equation $H_B \Psi=0$.
As for the AB-stacked bilayer, one obtains two distinct solutions, one with even and one with
odd parity, such that the zero-energy level
is orbitally two-fold degenerate, in addition to its usual four-fold spin-valley degeneracy and the orbital degeneracy described
by the flux density $n_B=eB/h$. The two zero-energy states, which are reminiscent of coherent states, 
are directly obtained from the secular equation:
\beqn\label{eq:states0}
\Psi_0 &=& 
\mathcal{N}_0 \cosh(\alpha^* a^{\dagger})\left(\begin{array}{c} 0 \\ |n=0\rangle \end{array}\right), 
\\
{\rm and} \qquad
\Psi_1 &=& 
\mathcal{N}_1 \frac{\sinh(\alpha^* a^{\dagger})}{\alpha^*}\left(\begin{array}{c} 0 \\ |n=0\rangle \end{array}\right),
\eeqn
in terms of the normalization factors $\mathcal{N}_{0/1}$. 
These states are generalizations of the zero-energy states $(0,|0\rangle)$ and $(0,|1\rangle)$ of the AB-stacked 
bilayer.\cite{MF06}. We also notice that we neglect any other potential lifts due to Zeeman effect and/or interactions.\cite{goerbig10} 
\begin{figure}[t]
\centering
{\label{bla2}\includegraphics[width=0.4\textwidth]{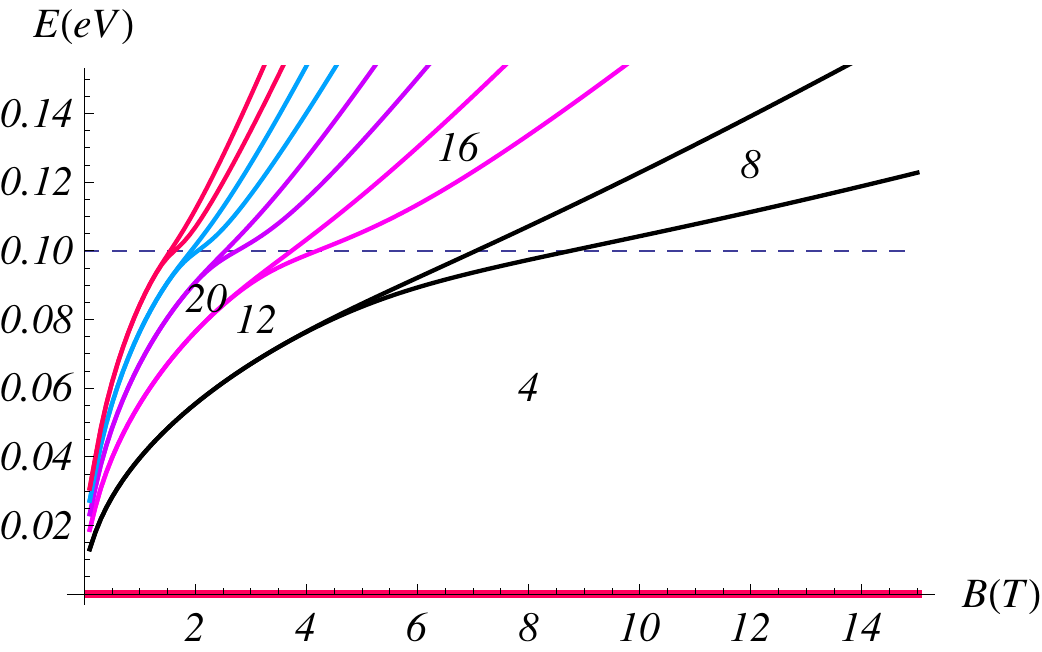}} 
\caption{(Color online) Landau levels of twisted bilayer graphene, obtained from the numerical solution of Eq. (\ref{eq:hamKK_B}). The characteristic
energy scale of the van-Hove singularity has been chosen as $|\alpha|^2\omega_C=(\Delta K)^2/8m=0.1$eV. 
The red line indicates the zero-energy level, and the numbers correspond to the filling factors in the gaps.
}
\label{fig02}
\end{figure}
The existence of those zero-mode states is independent of the strength of the magnetic field. This is to be compared to the other topological class of Dirac cones with opposite Berry phases where the degeneracy of the zero-mode is lifted.\cite{listDPmerg} This protection is solely determined by the topology of the model Hamiltonian since the band-structures are otherwise identical. Notice further that the scaling of the Landau levels relative to $\Delta \bK$ is different for the two topological classes.

\subsection{Full Landau-level spectrum}
\label{sec:nLL}

The full LL spectrum, which has been calculated numerically, is depicted in Fig. \ref{fig02}. In the regime of small magnetic fields,
the LLs with small index $n\ll n_C$, where $n_C$ denotes roughly the LL which crosses the van-Hove singularity, 
$\sqrt{n_CB}\sim v_F|\Delta K|^2/4\sqrt{2}\tilde{t}_{\perp}$, 
display the $\sqrt{Bn}$ behavior, which is the benchmark of massless Dirac fermions, as expected for the linear dispersion below 
the van-Hove singularity. Because of the two Dirac points, these LLs, as well as the zero-energy 
level discussed above, are twofold degenerate, in addition to the fourfold spin-valley degeneracy. The twofold degeneracy due to the
Dirac-point splitting in the twisted bilayer is lifted once the LLs approach and eventually cross the van-Hove singularity, 
$n\gtrsim n_C$, above which the LLs scale as $\sim B(n+1/2)$, as one would expect from the parabolic dispersion relation revealed 
by Hamiltonian (\ref{eq:hamKK}) at high energies. Notice, however, that as for AB-stacked bilayer graphene the effective two-band
approximation is no longer valid at higher energies (far beyond the van-Hove singularity) because of the presence of the remaining bands, which become visible there. 

The LL spectrum in Fig. \ref{fig02} allows one to understand the main features of a quantum Hall effect (QHE) in twisted bilayer 
graphene. The topologically protected zero-energy LL, with its altogether eightfold degeneracy, yields a QHE at filling 
factors $\nu=\pm 4$ (taking into account spin degeneracy). This feature is independent of the interlayer hopping strength $t_{\perp}$ and of the van-Hove singularity,
which is triggered by the twist angle $\theta$. For other LLs, the position of the van-Hove singularity determines their
degeneracy. If the LLs remain well below the singularity, $n\ll n_C$, they maintain their eightfold degeneracy, and one would therefore 
expect Hall plateaus at filling factors $\nu=\pm 4(2n+1)=\pm 4, \pm 12, \pm 20, ...$, but one would expect
additional plateaus at $\nu=\pm 8, \pm 16, ...$ for 
LLs with  $n\gtrsim n_C$. 
The experiment\cite{Lee10} indicates, even at rather low magnetic fields of $B\sim 5$T, only
the zero-energy LL is eightfold degenerate, whereas a plateau at $\nu=\pm 8$ has been observed. This stipulates that
the energy of the van-Hove singularity is as small as the first excited LL.  

\section{Conclusions}

In conclusion, we have investigated the topological band structure of twisted bilayer graphene, in the framework of a
symmetry analysis of the inter-layer hopping in the continuum limit. For small and moderate twist angles $\theta$, the two 
copies of the Dirac point (that are not related by time-reversal symmetry) are described by the same Berry phase, due
to the the symmetry of the inter-layer hopping term.
Therefore, they
belong to a different topological class than the (usual) two Dirac points, which are related by time-reversal symmetry.
In the presence of a quantizing magnetic field,
these particular topological properties yield a protected zero-energy LL with an eight-fold degeneracy that may be evidenced in 
quantum Hall transport measurements. 

\acknowledgements

This work was supported by the ANR project NANOSIM GRAPHENE under Grant No. ANR-09-NANO-016 and by the Ecole Doctorale de Physique de
la R\'egion Parisienne (ED 107), AHCN acknowledges partial financial support from the U.S. DOE under grant DE-FG02-08ER46512.
FG is supported by  by MICINN (Spain), grants FIS2008-00124 and CONSOLIDER CSD2007-00010.
We acknowledge fruitful discussions with Klaus von Klitzing, Nuno Peres, Jo\~ao Lopes dos Santos, and Jurgen Smet. 
AHCN thanks the Laboratoire de Physique des Solides for hospitality.


\begin{thebibliography}{99}
\bibitem{note}
The Hamiltonian (\ref{eq:ham1}) is similar to that of Ref. \onlinecite{MacDonald} up to a low-angle expansion.
\end{thebibliography}
\end{document}